\documentclass[a4paper,11pt]{article}
\usepackage{pos}
 \usepackage{graphicx}
 \usepackage{url}
 \usepackage{comment}

\newcommand\ResNbTwo{\overline{{\rm N}}_2}
\newcommand\NkLONkLLtimesNLOmt{{\rm (N}^{k}{\rm LO+N}^k{\rm LL)\otimes NLO}_{m_t}}
\newcommand\NkLOtimesNLOmt{{\rm N}^{k}{\rm LO \otimes NLO}_{m_t}}
\newcommand\NkLONkLL{{\rm N}^{k}{\rm LO+N}^k{\rm LL}}
\newcommand\NLOmt{{\rm NLO}_{m_t}}

 \title{Di-Higgs productions at N$^3$LO+N$^3$LL}
\author*[a]{A.H. Ajjath}
\author[a]{and Hua-Sheng Shao}
\affiliation[a]{Laboratoire de Physique Th\'eorique et Hautes Energies (LPTHE), UMR 7589, Sorbonne Universit\'e et CNRS, 4 place Jussieu, 75252 Paris Cedex 05, France}
\emailAdd{aabdulhameed@lpthe.jussieu.fr}
\emailAdd{huasheng.shao@lpthe.jussieu.fr}
\abstract{
In this talk, we discuss the soft-gluon resummation for a pair of Higgs bosons in the dominant gluon fusion channel to next-to-next-to-next-to-leading logarithmic (N$^3$LL) accuracy. Through the study, we achieve sub-percent level accuracy in the uncertainties from the residual renormalisation and factorisation scales, for both inclusive and differential mass distribution in the infinite top quark mass limit. We also combine these results with the full top mass dependent next-to-leading order calculations.}

\FullConference{16th International Symposium on Radiative Corrections: Applications of Quantum Field Theory to Phenomenology (
RADCOR2023)\\
28th May - 2nd June, 2023\\
Crieff, Scotland, UK\\}

\begin{document}
\maketitle

\section{Introduction}
 Over the past decade, there has been extensive studies on the Higgs boson and its fundamental properties such as the Higgs mass, spin and decay width. Notably, the Higgs interactions with $W,Z\gamma$, gluon and third generation charged fermions are compatible with the Standard Model (SM) within 10$\%$  accuracy\cite{ATLAS:2022vkf,CMS:2022dwd}. One of the pressing questions in the ongoing and upcoming Large Hadron Collider (LHC) experiments pertains to the self-interactions of the Higgs boson.
Understanding these self-interactions is crucial for unraveling the Higgs field potential and gaining deeper insights into the electroweak spontaneous symmetry breaking mechanism. 

The Higgs self interactions are accessible by studying the multiple-Higgs boson production from high-energy particle collisions.  
 Currently, the most viable production mode is the production of a pair of Higgs bosons, which provides direct access to Higgs trilinear coupling $\lambda_{hhh}$~\cite{DiMicco:2019ngk} and also offers indirect means to constrain the Higgs quartic coupling~\cite{Bizon:2018syu}. This discussions primarily focuses on the dominant production mode  $gg \rightarrow hh$, in which two incoming gluons emitted by the protons collides to yields a pair of Higgs bosons. 

In the standard model, the perturbative calculation of this process faces challenges due to the absence of tree-level interactions between the Higgs boson and gluons.
The leading-order (LO) cross section is derived from the square of the one-loop amplitude, prominently via the virtual top quark loop.
Extending these calculations beyond LO in perturbative QCD becomes increasingly complex due to intricate multi-loop, multi-scale Feynman diagrams. Thanks to several recent developments, the next-to-leading order (NLO) has been achieved in refs.~\cite{Borowka:2016ehy,Baglio:2018lrj}. Further improves have been obtained by soft gluon resummation~\cite{DeFlorian:2018eng} or matching to the parton showers~\cite{Heinrich:2017kxx}. These calculations represent the state-of-the-art results, incorporating the full top quark mass ($m_t$) dependence. However, they involves substantial theoretical uncertainties arising from large scale variations and the choice of top-quark mass scheme, which needs to be tackled in the future. 

In the absence of perturbative computations beyond NLO with the full $m_t$-dependence, another method for improving the cross sections is through "infinite top quark mass approximation", where $m_t\to \infty$. In this approximation, the Higgs boson interacts with the gluons through the effective vertices encoded in the effective Lagrangian.
This approach enables calculations beyond NLO, culminating in the next-to-next-to-next-to-leading order (N$^3$LO)~\cite{Chen:2019lzz, Chen:2019fhs}, thanks to the decades of theoretical physics efforts.
 The soft-gluon resummation effects before our work in~\cite{Ajjath:2022kpv} have been studied up to NNLO plus next-to-next-to-leading logarithmic (NNLL) accuracy in ref.~\cite{deFlorian:2015moa}. These results can be further improved by accommodating the finite $m_t$ corrections by rescaling them with the known full-$m_t$ dependent cross sections ~\cite{Maltoni:2014eza,Frederix:2014hta,Grazzini:2018bsd}. This talk discusses the work presented in~\cite{Ajjath:2022kpv}, where we further improve the perturbative QCD calculations for $gg\to hh$ by employing threshold resummation up to next-to-next-to-next-to-leading logarithmic (N$^3$LL) accuracy in the infinite top quark mass approximation. We also incorporate the finite top mass corrections in our results at N$^3$LO+N$^3$LL in the $m_t \rightarrow \infty $ limit with the known full $m_t$-dependent NLO calculations.

\section{Theoretical aspects\label{sec:theoryframe}}

In QCD improved parton model, the differential invariant mass distributions for a heavy colourless final states with invariant mass $M^2$ produced in hadron collisions takes the form of a convoluted integral:
\begin{equation}\label{eq:partonmodel}
    M^2 \dfrac{d}{d M^2}\sigma_{h_1h_2}(s,M^2) = \tau \sum_{a,b=q,\bar q,g} \int_\tau^1 \frac{dz}{z} \phi_{ab}\left(\frac{\tau}{z},\mu_F^2\right)~ \Delta_{ab}  \left(z,M^2,\mu_F^2\right)\,,
\end{equation}
with $s$ denotes the square of the hadronic center-of-mass energy and $\tau=\dfrac{M^2}{s}$ is the dimensionless hadronic scaling variable. Similarly, the partonic scaling variable $z=\dfrac{M^2}{\hat s}$ with $\hat s$ being the partonic centre of mass energy.
  The partonic luminosity $\phi_{ab}$ is defined as:
\begin{equation}\label{eq:partonlum}
    \phi_{ab}(x,\mu_F^2) \equiv \int_x^1 \dfrac{dy}{y} f_{a/h_1}(y,\mu_F^2) f_{b/h_2}\left(\dfrac{x}{y},\mu_F^2\right).
\end{equation}
with the factorisation scale $\mu_F$ and the incoming parton distribution functions $f_{c/h_i}$. The sum $\sum_{a,b=q,\bar q,g}$ refers to the species of incoming partonic states. The $\Delta_{ab}  \left(z,M^2,\mu_F^2\right)$ is the perturbatively calculable coefficient functions, which can be decomposed according to their singular behaviour as follows,
\begin{align}
\label{eq:bBH-PartsOfDelta}
 &{\Delta}_{ab}(z, M^{2}, \mu_F^2) = {\Delta}^{
  \text{SV}}_{ab} (z, M^{2},  \mu_F^2) + {\Delta}^{
  \text{Reg}}_{ab} (z, M^{2},  \mu_F^2) \,.
\end{align}
Both the terms in \eqref{eq:bBH-PartsOfDelta} satisfy perturbatively expansion in the renormalised strong coupling constant.
In the aforementioned decomposition, the term $\Delta^{
   \text{SV}}_{ab} (z, M^{2},  \mu_F^2) $ involves only the singular terms when $z=1$ with the distributions of the following forms:
   \begin{equation}
      \left\{ \delta(1-z), {\cal D}_i(z) \equiv  \left[\frac{\ln^i{(1-z)}}{1-z}\right]_+ \right\}\,,
   \end{equation}
where $\delta(1-z)$ is the Dirac delta function and ${\cal D}_i(z)$ is the plus distribution.
The superscript $\text{SV}$ denotes the soft-virtual part, accounting for the real emission corrections at the threshold limit and the pure virtual corrections. The remaining regular term $\Delta^{\text{Reg}}_{ab} (z, M^{2}, \mu_F^2)$ involves either $\ln^i{(1-z)}$ or the polynomials of $(1-z)$. 
Although the logarithms $\ln^i{(1-z)}$ could also be large, they are subdominant by a factor of $(1-z)$ compared to the soft-virtual contribution.

The soft-virtual contribution can be expressed in terms of hard function $H_{ab}$ and the soft-collinear function $S_{\Gamma,ab}$:
\begin{align}\label{eq:partonicxsec}
    \Delta^{\text{SV}}_{ab}(z,M^2,\mu_F^2) =& H_{ab}(M^2,\mu_R^2) ~\delta(1-z)\otimes S_{\Gamma,ab}(z,M^2,\mu_F^2,\mu_R^2)\,.
\end{align}
Here $\mu_R$ stands for the renormalisation scale of the strong coupling constant $\alpha_s$ .
 The hard function $H_{ab}$ is obtained from the infrared (IR) subtracted loop corrections, and the soft-collinear function $S_{\Gamma,ab}$ represents the real corrections and the contribution of the Altarelli Parisi splitting kernels in the soft limit. Both $H_{ab}$ and $S_{\Gamma,ab}$ are free of IR poles. For more details on their structure and for explicit results, we refer the readers to  refs.~\cite{Ravindran:2005vv, Ravindran:2006cg, Ahmed:2020nci,Ajjath:2022kpv,Ahmed:2020amh,Ajjath:2019neu}.

In the threshold limit, where the incoming partonic centre of mass energy is close to the invariant mass of the Higgs pairs, the soft-virtual corrections dominate by the large logarithms of the form $\ln(1-z)$. This potentially spoils the perturbative convergence of the coefficient function, if we truncate them at a fixed order in the strong coupling constant. Therefore we need the threshold resummation~\cite{Sterman:1986aj,Catani:1989ne}, which provides an alternative perturbative expansion. 

The $z$-space coefficient $\Delta^{\text{SV}}_{ab}\left(z,M^2,\mu_F^2\right)$ involves integral convolutions, which is convenient to express in the Mellin $N$-space, where all the convolutions turn into normal products. The Mellin transform relates the partonic coefficient function in the $z$ and $N$-space by
\begin{equation}
    \Delta_{ab}^{\rm res}(N,M^2,\mu_F^2) = \int_0^1 dz ~z^{N-1}\Delta^{\text{SV}}_{ab}\left(z,M^2,\mu_F^2\right)\,,
\end{equation}
with the threshold limit in the Mellin space represented by the large $N$ limit. Performing the resummation in the Mellin space resolves the issues of large logarithms by reorganising the series in the resummation parameter $(\omega \equiv 2 \beta_0 a_s(\mu_R^2) \ln N )$ at everty order. The SV resummation for the coefficient function in the Mellin space reads:
\begin{align}\label{DeltaN}
    \Delta_{ab}^{\rm res}(N,M^2,\mu_F^2) &= g_{0,ab}(M^2,\mu_F^2,\mu_R^2)\times \nonumber\\
    &~\exp \left(\tilde{C}_{0,ab} (a_s(\mu_R^2)) + g_{1,ab} (\omega)\ln N + \sum_{k=2}{a_s^{k-2}(\mu_R^2)~g_{k,ab} (\omega)}\right)\,.
\end{align}
where $g_{0,ab}$ results from the hard function and Altarelli Parisi splitting kernels in the Mellin space, $\tilde{C}_{0,ab} $ denotes the non-logarithmic part of soft function and the $N$-dependent terms in the exponent arises from soft-logarithmic contributions. For processes with colourless final states, the resummation coefficients $g_{k,ab}(\omega)$ and $\tilde{C}_{0,ab}$ are universal. They only depends on the nature of the incoming particles. The exponential form in eq.\eqref{DeltaN} reorganises the perturbative series and enables the resummation of the particular logarithmic terms to all orders in the strong coupling constant.

When the resummation calculations match to the fixed-order computations, we need to subtract the double counting between the two. At the $\NkLONkLL$ accuracy, the double counting is just the N$^k$LO soft-virtual contributions encoded in the partonic coefficient
function. However, up to a given logarithmic accuracy, the right-hand side of the resummation formalism eq.~(\ref{DeltaN}) is not uniquely defined. One has the freedom to choose which $N$-independent piece should be absorbed into the exponent. 
 This essentially leads to ambiguities in the threshold resummation formalism at a given logarithmic accuracy. Such resummation ambiguities have also beeen studied in  ref.~\cite{Ajjath:2020rci}. For more details on the resummation ambiguities in our study, we refer the reader to section~2.4 in~\cite{Ajjath:2022kpv}.

\section{Results\label{sec:results}}
In our numerical calculations, we adopt the following setup: the Higgs mass is set at $m_h=125$ GeV, the vacuum expectation value of the Higgs field at $v=246.2$ GeV, the top-quark pole mass $m_t=173$ GeV and we employ the PDF {\tt PDF4LHC15\_nnlo\_30} provided by {\sc\small LHAPDF6}, along with its renormalisation group running for strong coupling constant. We choose the default central scale as $\mu_0=m_{hh}/2$, where $m_{hh}$ is the invariant mass of the Higgs pair. To assess the scale uncertainties, we utilise the standard $7$-point scale variation approach.

Our initial focus centers on the study of resummation calculations for various resummation schemes pertaining to the inclusive cross section. The ambiguity in resummation schemes arises from the treatment of the non-logarithmic part of the soft function, either within the exponent or outside in eq.\eqref{DeltaN}. We observe that the scheme dependence reduces from NLO+NLL to N$^3$LO+N$^3$LL. Notably, we find the least scale uncertainties in the so-called $\ResNbTwo$ scheme, where we resum the complete soft function contributions, including the non-logarithmic part. Beyond the scale uncertainties, it is also worth to note that the central values of N$^k$LO+N$^k$LL in the $\ResNbTwo$ scheme are remarkably close to N$^{k+1}$LO, than in other schemes. Consequently, we adhere to the $\ResNbTwo$ scheme for the subsequent discussions.
\begin{figure*}[h]
\centering
\includegraphics[width=0.9\textwidth]{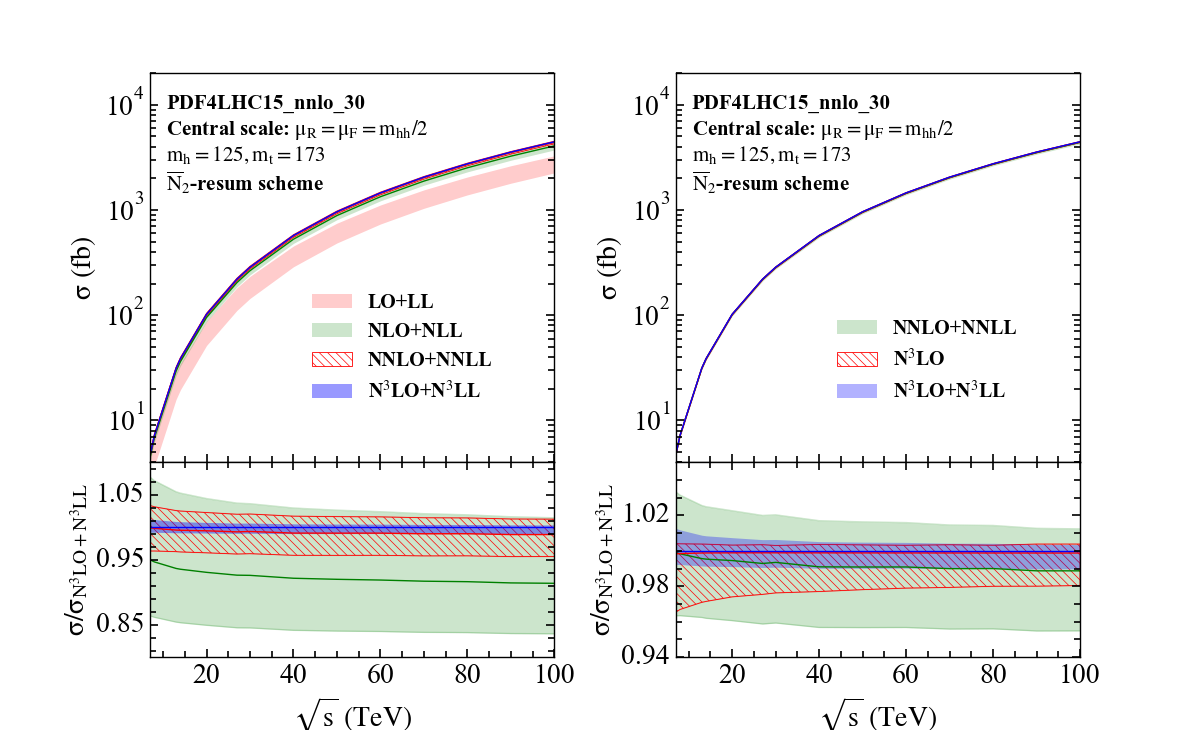}
\caption{Inclusive cross sections with respect to $\sqrt{s}$ from LO+LL to N$^3$LO+N$^3$LL (left) and comparison study of the best result N$^3$LO+N$^3$LL against N$^3$LO and NNLO+NNLL (right), with the 7-point scale uncertainties. The lower panels depicts the ratios over the central value of N$^3$LO+N$^3$LL.}
\label{fig:Bandplot_best_NExp}
\end{figure*}

We illustrate the inclusive cross sections upto N$^3$LO+N$^3$LL in figure~\ref{fig:Bandplot_best_NExp} over a broad energy range, spanning from $\sqrt{s}=7$ TeV to $\sqrt{s}=100$ TeV.
 We find that the N$^3$LO+N$^3$LL exhibits only minimal enhancements compared to N$^3$LO, affirming the conclusion arrived in \cite{Chen:2019fhs} that the asymptotic convergence in the strong coupling constant has been achieved at N$^3$LO. The scale uncertainties of N$^3$LO+N$^3$LL are a factor of $2$ smaller than the N$^3$LO and nearly one-fourth of those in NNLO+NNLL, reaching to the sub-percent level. Notably, the N$^k$LO+N$^k$LL error bands, representing the $7$-point scale variations, are fully encompassed within the previous order error bands when $k\geq 2$. At $13$~TeV, the NLO+NLL surpasses LO+LL by $40\%$, which is further enhanced at NNLO+NNLL by $6\%$ and N$^3$LO+N$^3$LL by $0.4\%$. A comparison study of the best results for NNLO+NNLL, N$^3$LO, and N$^3$LO+N$^3$LL can be found in the right panel of figure~\ref{fig:Bandplot_best_NExp}, where the N$^3$LO error band is entirely encompassed within the NNLO+NNLL error band, and the N$^3$LO+N$^3$LL error band largely resides within the N$^3$LO. This again points to the good asymptotic perturbative convergence at N$^3$LO.

\begin{figure*}[htbp]
\centering
\includegraphics[width=0.41\textwidth]{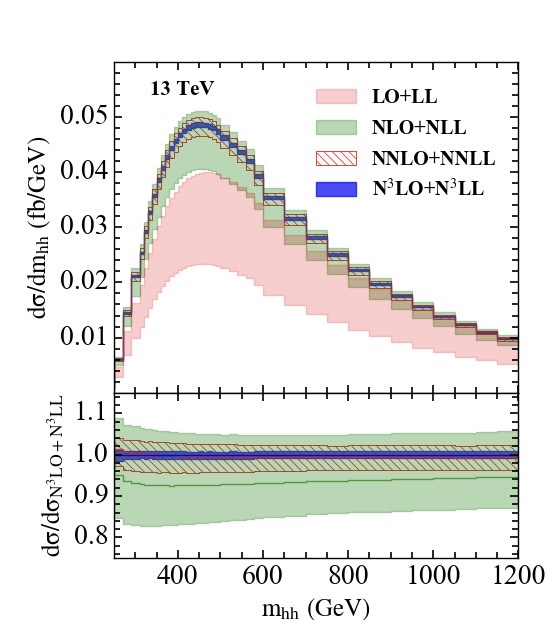}
\includegraphics[width=0.43\textwidth]{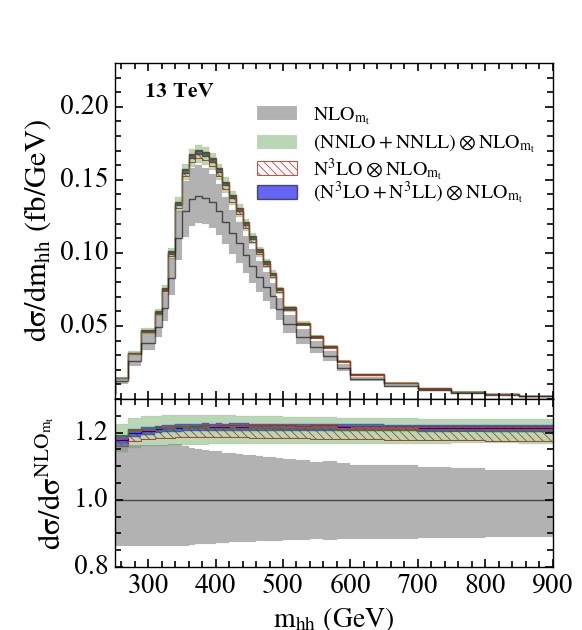}
\\
\caption{The invariant mass distribution in the $m_t\rightarrow \infty$ limit (left) and with finite top quark mass corrections (right) from LO+LL to N$^3$LO+N$^3$LL at  $\sqrt{s}=13$ TeV. The error bands represent the 7-point scale variations with the central scale $\mu_0=\frac{m_{hh}}{2}$.}
\label{fig:xs_vs_mhh_large_mt1}
\end{figure*}

We also display the invariant mass distribution, $\frac{d\sigma}{dm_{hh}}$, for Higgs boson pairs upto the
 N$^3$LO accuracy at 13 TeV centre of mass energy, in the $m_t\to\infty$ approximation (as depicted in the left panel of figure~\ref{fig:xs_vs_mhh_large_mt1}). The figure illustrates the profound impact of incorporating higher-order QCD radiative corrections, resulting in a stabilisation of perturbative calculations for the invariant mass differential distributions. Additionally, the inclusion of these higher-order corrections considerably reduces the scale uncertainties, similar to the  case of the inclusive cross section.

Further we combine our results in the infinite top quark mass approximations with the NLO full $m_t$-dependent results (denoted as ``$\NLOmt$''). The inclusion of finite top quark mass corrections is crucial in the case of $gg\to hh$ for realistic phenomenological applications. The $\NLOmt$ results are directly computed using the publicly available code~\cite{Heinrich:2017kxx,Heinrich:2019bkc}. To improve the (differential) cross section, we multiply the (differential) QCD $K$ factors obtained from the $m_t\to \infty$ calculations with $\NLOmt$, a combination referred to as $\NkLONkLLtimesNLOmt$ in ref~\cite{Ajjath:2022kpv}. The cross section for each scale choice is evaluated following the methodology outlined in eq.(3.4) in ref.~\cite{Chen:2019fhs}. The relative scale errors in $\NkLONkLLtimesNLOmt$ and $\NkLOtimesNLOmt$ are identical to those in N$^k$LO+N$^k$LL and N$^k$LO in the infinite top quark mass approximation. Although we have only considered $\NLOmt$, it is worth noting that our calculations facilitate obtaining the most advanced theoretical predictions. They will still be valuable even after the full $m_t$-dependent NNLO results have become available. In the right panel of figure~\ref{fig:xs_vs_mhh_large_mt1}, we present predictions for the invariant mass distributions, accounting for the NLO full top-quark mass dependence.

Besides the scale uncertainties we have discussed so far, other large uncertainties can arise from the top quark mass scheme choice and the missing finite top quark corrections beyond NLO. Since our work focuses on the infinite top quark mass approximation, the inclusion of higher-order $\alpha_s$ corrections is unlikely to alleviate such uncertainties.
\section{Summary\label{sec:conclusions}}
In this talk we have discussed our recent improvements regarding the (differential) cross section for the production of a pair of Higgs bosons via gluon-gluon fusion. We have incorporated the threshold resummation up to N$^3$LO+N$^3$LL accuracy within the infinite top quark mass approximation. Notably, our results have yielded the residual renormalisation and factorisation scale uncertainties at a remarkable sub-percent level, which are a factor of $2$ smaller than N$^3$LO and four times smaller than NNLO+NNLL results. Moreover, the central values are very close to the N$^3$LO results, affirming the conclusion in~\cite{Chen:2019fhs}
that the asymptotic perturbative convergence in the strong coupling constant has been achieved at N$^3$LO. 
Our results are finally combined with the full top quark mass dependent calculations at NLO QCD for the phenomenological applications.


\bibliographystyle{JHEP}
\bibliography{main}
\end{document}